\documentclass[letterpaper, 10 pt, conference]{IEEEtran}
\IEEEoverridecommandlockouts
\usepackage{cite}
\usepackage{amsmath,amssymb,amsfonts}
\usepackage{algorithmic}
\usepackage{graphicx}
\usepackage{textcomp}
\usepackage{xcolor}
\def\BibTeX{{\rm B\kern-.05em{\sc i\kern-.025em b}\kern-.08em
    T\kern-.1667em\lower.7ex\hbox{E}\kern-.125emX}}
    
\newcommand\scalemath[2]{\scalebox{#1}{\mbox{\ensuremath{\displaystyle #2}}}}
\usepackage{float}

\usepackage[subtle,title,leading=tight,paragraphs=tight,tracking=tight,wordspacing=tight]{savetrees}
\usepackage{microtype} 
\usepackage[font=footnotesize]{caption}
\begin{document}
\title{\LARGE \bf Nonlinear Model Based Guidance with Deep Learning Based Target Trajectory Prediction  Against Aerial Agile Attack Patterns}
% \\
% {\footnotesize \textsuperscript{*}Note: This is the author version of the manuscript of the same name accepted in the proceedings of the 2021 American Control Conference (ACC 2021)}}
\author{A. Sadik Satir$^{1}$, Umut Demir$^{2}$, Gulay Goktas Sever$^{3}$, N. Kemal Ure$^{4}$
% <-this % stops a space
%\thanks{*This work was supported by Aselsan}% <-this % stops a space

\thanks{This is the author version of the manuscript of the same name accepted 
in the proceedings of the 2021 American Control Conference (ACC 2021).}
\thanks{$^{1}$MSc Student, Department of Aeronautics and Astronautics, Istanbul Technical University, Istanbul, Turkey.
        {\tt\small sadiksatir@gmail.com}}
\thanks{$^{2}$MSc Student, Department of Aeronautics and Astronautics, Istanbul Technical University, Istanbul, Turkey.
        {\tt\small demiru@itu.edu.tr}}
\thanks{$^{3}$MSc Student, Department of Defense Technologies, Istanbul Technical University, Istanbul, Turkey.
        {\tt\small sever17@itu.edu.tr}}

\thanks{$^{4}$ ITU Artificial Intelligence and Data Science Application and Research Center, Associate Professor, Department of Aeronautics and Astronautics, Istanbul Technical University, Istanbul, Turkey.
        {\tt\small ure@itu.edu.tr}}
}
\date{}
\maketitle

\begin{abstract}
In this work, we propose a novel missile guidance algorithm that combines deep learning based trajectory prediction with nonlinear model predictive control. Although missile guidance and threat interception is a well-studied problem, existing algorithms' performance degrade significantly when the target is pulling high acceleration attack maneuvers while rapidly changing its direction. We argue that since most threats execute similar attack maneuvers, these nonlinear trajectory patterns can be processed with modern machine learning methods to build high accuracy trajectory prediction algorithms. We train a long short-term memory network (LSTM) based on a class of simulated structured agile attack patterns, then combine this predictor with quadratic programming based nonlinear model predictive control (NMPC). Our method, named nonlinear model based predictive control with target acceleration predictions (NMPC-TAP), significantly outperforms compared approaches in terms of miss distance, for the scenarios where the target/threat is executing agile maneuvers.
\end{abstract}

% \begin{IEEEkeywords}
% component, formatting, style, styling, insert
% \end{IEEEkeywords}

\section{INTRODUCTION}
%%%%%%%%%%%%% Intro with history %%%%%%%%%%%%%%%%%
Since the World War II, missile interception guidance has been one of the most important problems in missile studies. Several guidance laws are introduced over the years. One of the most popular and commonly used guidance law is Proportional Navigation (PN). PN can be implemented easily and can guarantee to hit targets under certain limitations such as non-maneuvering or weakly maneuvering targets \cite{zarchan2012tactical}. Garber proposed Augmented Proportional navigation (APN) to overcome these limitations\cite{garber1968optimum}. APN is developed by further optimizing the PN guidance. Thereafter, several other guidance laws are designed based on APN \cite{ghosh2013capturability} \cite{ghosh2014capturability}, but none of these approaches are capable of explicitly predicting the future trajectory of the target. It is shown that the success of guidance algorithms can be improved with prediction of target maneuvers. Moreover, if target performs agile maneuvers and the distance between target and interceptor is small, predicting the future states of the target can be used to increase the hit performance \cite{zhurbal2011effect}. That being said, intercepting highly agile targets is still an open problem, since the predictive capabilities of existing algorithms are usually limited to targets that execute standard (non-agile) maneuvers. The main objective of this paper is to utilize recent advances in machine learning and predictive guidance to push the state of the art performance in intercepting agile targets.

%%%%%%%%%%%%%%%%%%%%% Literature Review %%%%%%%%%%%%%%%%%%%%%%%%%
\subsection{Previous Work}
In recent years, there has been a great deal of effort in developing predictive guidance laws by employing knowledge/prediction of target trajectory and/or dynamics \cite{prabhakar2013predictive,ure2016predictive,akcal2017predictive}.
Best and Norton\cite{best2000predictive} proposed a predictive guidance law inspired by model predictive control (MPC). The guidance law is based on predicting the probability density function(PDF) of target position at interception,
Shima et al. \cite{shima2002efficient}  presented the multiple-model estimator in case the target performs evasive maneuver characterized by a random switching time.
Dionne et al. \cite{dionne2006predictive} extends the Best and Norton \cite{best2000predictive} and presents a new terminal guidance law which uses target position that estimated a in terms of a PDF in the terminal constraint.
% \cite{dionne2007predictive} 
% Prabhakar et al. \cite{prabhakar2013predictive} proposed a predictive explicit guidance scheme for ballistic missiles. The missile model is used to predict its impact point deviations at every cycle. 
Nobahari \cite{nobahari2016heuristic} introduced a heuristic predictive line-of-sight (LOS) guidance law based on a dynamic optimization algorithm which determines the acceleration commands.
Ure \cite{ure2016predictive} proposed a stochastic predictive guidance and control methodology for interception of agile targets.
Akcal \cite{akcal2017predictive} presents a predictive guidance which uses a recursive least squares (RLS) to estimate the possible target positions and nonlinear programming is employed for selecting the optimal action.
% Yang \cite{yang2018predictive} A new aircraft guidance law using stochastic predictive control aims at maneuverable targets is proposed in this paper
Eren et al.\cite{MPCsurvey_2017_AIAA}  extensively investigated potential and vast applications of MPC in aerospace guidance field in the survey paper.
Li et al.\cite{2015PDNN} introduced a NMPC scheme in case of target acceleration is unknown for missile interception based on the primal–dual neural-network(PDNN) optimization. NMPC formulated as quadratic programming (QP) problem which needs to be solved at each time step  and PDNN optimization implementation was shown.
Integrated control and guidance of missiles based on NMPC proposed by Bachitar \cite{bachtiar2017nonlinear}. A multiobjective offline tuning of NMPC introduced which considered control performance and implementation cost based on required computational capacity.
% Chai et al. \cite{2019model} (integrated missile guidance and control presented based on nonlinear RHPC, nonlinear programming sensitivity based algorithm is used and embedded in the optimization framework.)
Kumar et al. \cite{kumar2020nonlinear} presented an integrated missile guidance and control using NMPC for missile-on-missile interception.To genarate an optimal control command constrained nonlinear problem solved with a sequential quadratic programming (SQP) algorithm.
Bhattacherjee \cite{2020NMPC} derived a NMPC  based missile guidance algorithm for target interception. 

\subsection{Contributions}
In this paper, nonlinear model predictive control with target acceleration predictions (NMPC-TAP) method is proposed by exploiting target trajectory predictions generated by a deep neural network. The main idea behind using deep neural networks for this purpose stems from the fact that a large array of past target trajectories are available from operational heuristics and domain expertise in air defense systems \cite{lotter2017design}. By using these heuristic, synthetic agile target trajectories can be fed to a time series prediction network (long short-term memory networks in our case) to predict target trajectories. We show that using such networks for agile maneuvering targets is indeed more efficient compared to alternative approaches.
Proposed method needs instantaneous and future target acceleration predictions along the prediction horizon of NMPC. Then a QP is solved at each time step by employing these predictions.
In PDNN \cite{2015PDNN} and  QP \cite{2020NMPC} based optimization approaches,control inputs and bounded unknown target accelerations are combined in a decision variable vector.
The QP problem in NMPC-TAP is obtained by leaving the control input as only decision variable, since the target accelerations are predicted. The NMPC-TAP is validated by simulations for agile maneuvers. It is shown that NMPC-TAP yields significantly superior performance compared to PN, APN and NMPC with unknown target acceleration in agile maneuvering target interception scenarios. 

The paper is organized is follows; the Section 2 gives a brief background on the interception geometry , the Section 3 provides the details of the NMPC based predictive guidance algorithms, the Section 4 describes the structure of the LSTM and finally the Section 5 gives simulation results for various guidance laws.

%%%%%%%%%%%%%%%%%%%%%%%%%%%%%%%%%%%%%%%%%%%%%%%%%%%%%%%%%%%%%%%%%%%%%%%%%%%%%%%%
%%%%%%%%%%%%%%%%%%%%%%%%%%%% BACKGROUND %%%%%%%%%%%%%%%%%%%%%%%%%%%%%%%%%%%%%%%%%
\section{INTERCEPTION GEOMETRY}
\begin{figure}[h!]
    \centering
    \includegraphics[scale=0.18]{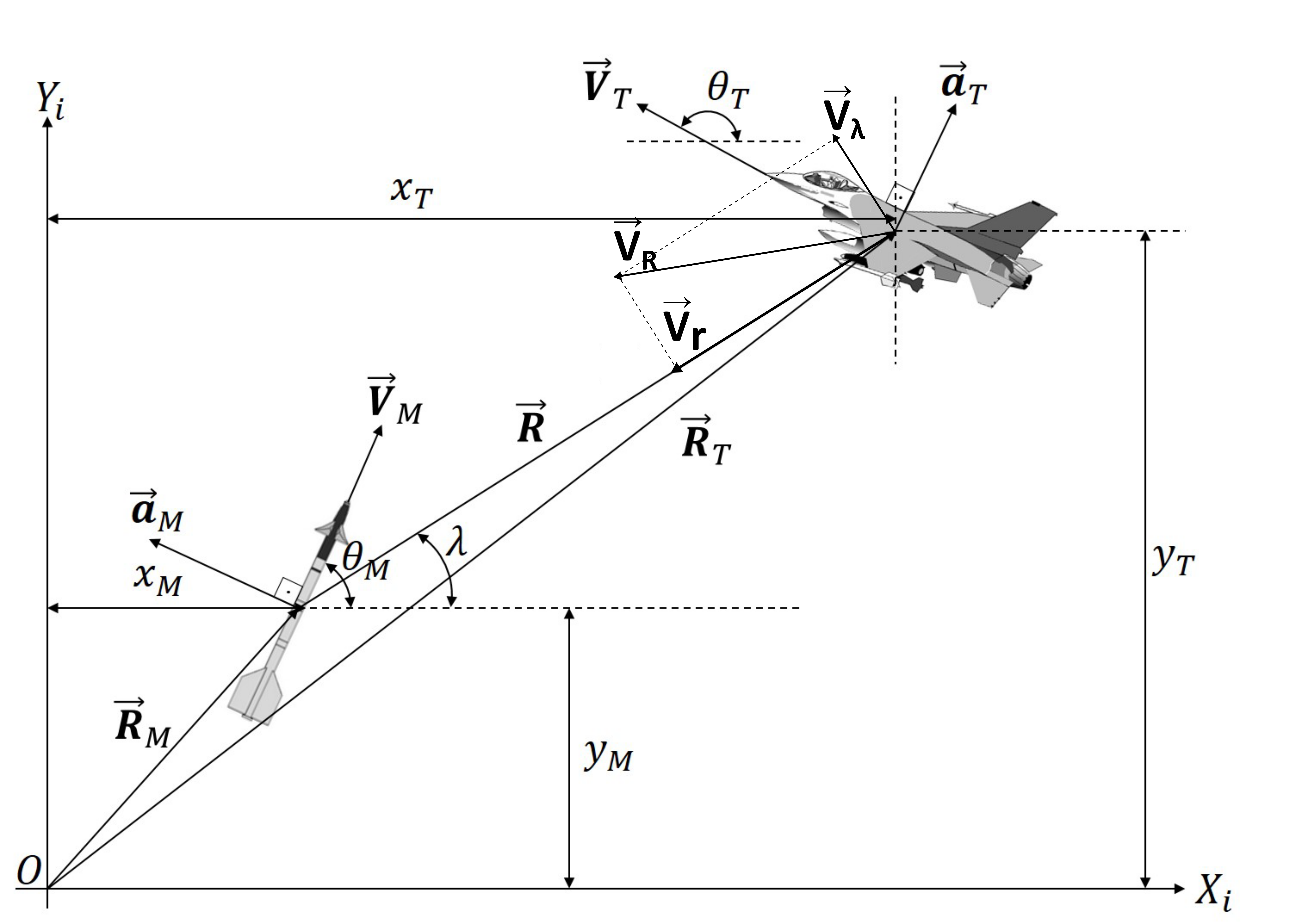}
    \caption{The missile-target engagement geometry.}
    \label{fig:dynamics}
\end{figure}
We consider a two-dimensional intercept model which is depicted in Figure 1.Assuming an inertial frame defined by $X_i$ and $Y_i$ axes together with corresponding unit vectors $\overrightarrow{i}$ and  $\overrightarrow{j}$. $x_M$; $y_M$; $x_T$ ; and $y_T$ are the Cartesian coordinates of the missile and the target while $\overrightarrow{R_M}$;$\overrightarrow{R_T}$; $\overrightarrow{V_M}$ and $\overrightarrow{V_T}$ represent the position and velocity vectors of the missile and target.In addition to the previous definitions, missile target range vector $\overrightarrow{R}$ can be rewritten in polar coordinates as $R=(r,\lambda)$. $r$ is the range along LOS, and $\lambda$ is the LOS angle. $V_R$ is the first time derivative of $R$ ; $V_T$ and $a_T$ are target velocity and acceleration, respectively; and $V_M$ and $a_M$ are missile velocity and acceleration, respectively.The interception geometry is represented in polar coordinate system in \cite{2001model}\cite{2015PDNN}.
\begin{align*}
    V_r=&V_T\cos (\theta _{T}-\lambda)-V_M\cos (\theta _{M}-\lambda)\\
    V_\lambda=&V_T\sin (\theta _{T}-\lambda)-V_M\sin (\theta _{M}-\lambda)\\
    a_{Tr}=&a_T\sin (\theta _{T}-\lambda)\\
    a_{T\lambda }=&a_T\cos (\theta _{T}-\lambda)
\end{align*}
$V_{\lambda}$ is a transversal component of relative velocity rotating with LOS, $a_{Tr}$ and $a_{T\lambda}$ can be described as the projection components of the target acceleration.
Let us consider the state variable vector $x=[r,V_r,\lambda,V_{\lambda}]^T$. Then, the following state model of missile-target interception model can be obtained in polar coordinates as follows:
\begin{equation}
\dot x=f(x)+g(x)u + d(x)w, \label{eq:continuous ss1}
\end{equation}
where $x \in\Re^{n_x}$ is a state vector, $w=[a_{Tr},a_{T\lambda}]^T\in \Re^2$ is a vector which represents target acceleration  in polar coordinates, and $u=a_M \in \Re$ is a control input.
The control input is subjected to an acceleration limit, which is expressed in Eq. \ref{eq:limit}.
\begin{equation}
\label{eq:limit}
    -u_{max} \leq u \leq u_{max}.
\end{equation}
It is well known that MPC usually requires difference equations in discrete-time. Difference equations can be easily obtained by applying Euler discretization on Eq.(\ref{eq:continuous ss1}):
\begin{equation}
x(k+1)=f_d(x(k))+g_d(x(k))u(k)+d_d(x(k)) w(k) \label{eq:discrete ss}
\end{equation}
where
\begin{align*}
f_d(x(k))=&\begin{bmatrix} x_1(k)\\x_2(k)\\x_3(k)\\x_4(k) \end{bmatrix} + \Delta(t)\begin{bmatrix}x_2(k)\\x_4^2(k)/x_1(k)\\x_4(k)/x_1(k)\\-x_2(k)x_4(k)/x_1(k) \end{bmatrix}\\
g_d(x(k))=&\Delta(t)\begin{bmatrix}0\\sin(\theta_M-x_3(k))\\0\\-\cos(\theta_M-x_3(k)) \end{bmatrix}, d_d(x(k))=\Delta(t)\begin{bmatrix}0\\a_{Tr}\\0\\a_{T\lambda} \end{bmatrix}
\end{align*}

%%%%%%%%%%%%%%%%%%%%%%% Guidance Algorithms %%%%%%%%%%%%%%%%
\section{PREDICTIVE GUIDANCE ALGORITHMS WITH CONSTRAINED OPTIMIZATION}
In this section QP based formulations for predictive guidance will be derived for both predicted and unknown target accelerations. 

\subsection{NMPC with Predicted Target Accelerations}
The NMPC-TAP method is described in this section. \\
The reference predictive form \cite{2006receding} can be expressed by using Eq. (\ref{eq:discrete ss}) as
\begin{align}
     x(k+j|k)=&f_d(x(k+j-1|k))+\\
     &g_d(x(k+j-1|k))u(k+j-1|k)+\nonumber\\
     &d_d(x(k+j-1|k))w(k+j-1|k)\nonumber
\end{align}
where
\begin{equation*}
    u(k+j-1|k)=u(k+j-2|k)+\Delta u(k+j-1|k)
\end{equation*}
where $(k+j|k)$ means that the current time-step is $k$ and the distance from the current time-step is $j$ \cite{2006receding}. The standard prediction form is given by
\begin{equation}
X_k=F_k + G_k\Delta U_k+g_k + d_k \label{eq:prediction ss}
\end{equation}
with
\begin{equation*}
\begin{array}{l}
      X_k=[x(k+1|k),\quad x(k+2|k),...,x(k+N_p|k)]^T \\
      U_k=[u(k|k),\quad u(k+1|k),...,u(k+N_p-1|k)]^T\\
      \Delta U_k=[\Delta u(k|k),\quad \Delta u(k+1|k),...,\Delta u(k+N_p-1|k)]^T\\
\end{array}
\end{equation*}

\begin{align*} 
F_k=&\left[ \begin{array}{c} f_d(x(k|k)) \\ f_d(x(k+1|k)) \\ \vdots \\ f_d(x(k+N_p-1|k)) \end{array}\right]
\\
G_k=& \scalemath{0.9}{\left[  \begin{array}{ccccccc} g_d(x(k|k))&\quad \dots &\quad 0 \\ g_d(x(k+1|k)) &\quad \dots &\quad 0 \\ \vdots &\quad \ddots &\quad \vdots \\ g_d(x(k+N_p-1)|k) &\quad \dots &\quad g_d(x(k+N_p-1)|k) \end{array}\right ]} \\ 
g_k=&\left [ \begin{array}{c} g_d(x(k|k)u(k-1)) \\ g_d(x(k+1|k))u(k-1) \\ \vdots \\ g_d(x(k+N_p-1|k-1))u(k-1) \end{array}\right ] \\ 
d_k=&\left [ \begin{array}{c}d_d(x(k|k))w(k|k)\\ \vdots\\  d_d(x(k+N_p-1|k))w(k+N_p-1|k)\end{array}\right ] . \end{align*}

where $u(k-1|k)$ is the control input of the last time-step, $N_p$,$N_c$ are the prediction and control horizons, respectively, $X_k \in\Re^{n_xN_p\times 1}$, $F_k \in\Re^{n_xN_p\times 1}$, $G_k \in\Re^{n_xN_p\times N_c}$, $\Delta U_k \in\Re^{N_c\times 1}$,$g_k \in\Re^{n_xN_p\times 1}$ and $D_k \in\Re^{n_xN_p\times 2}$.
Calculating the matrices $F_k$, $G_k$, $g_k$ is challenging, since the $\Delta U_k$ is unknown until optimization problem is solved. The problem is tackled by 
utilizing $\Delta U_k$ belongs to previous time step to calculate these matrices.  
\begin{equation}
    \begin{array}{ccc}
         \Delta u(k|k)&\triangleq& \Delta u((k-1)+1|k-1) \\
         \Delta u(k+1|k)&\triangleq& \Delta u((k-1)+2|k-1)\\
         \vdots&&\vdots\\
         \Delta u(k+N_p-1|k)&\triangleq& \Delta u((k-1)+N_p|k-1)\\
         \Delta u(k+N_p|k)&\triangleq& 0
    \end{array}
\end{equation}

We can define a quadratic cost function as:
\begin{equation}
    J=X_k^TQX_k\!+\Delta U_k^TR\Delta U_k \label{eq:cost}
\end{equation}
where $Q \in\Re^{n_xN_p\times n_xN_p}$ and $R \in\Re^{N_c\times N_c}$.
Then, the performance objective function can be rewritten by substituting Eq.(\ref{eq:prediction ss}) into Eq. (\ref{eq:cost}) 
\begin{align}
         J=&(F_k + G_k\Delta U_k+g_k + d_k)^TQ(F_k + G_k\Delta U_k+g_k+
         d_k)\!+\nonumber\\ 
         &\Delta U_k^TR\Delta U_k \label{eq:cost2}
\end{align}
The control input magnitude Eq. \ref{eq:limit} and its rate is constrained as:
\begin{equation}
    U_{\min } \leq  U_k \leq U_{\max }= U_{\min } \leq  U_{k-1}+I_{lt}\Delta U_k \leq U_{\max }
\end{equation}
where $U_{k-1}\in\Re^{N_c\times 1}$ and $I_{lt} \in\Re^{N_c\times N_c}$ are given as following
\begin{align}
    U_{k-1}=u(k-1|k)1_{N_c}\nonumber\\
    I_{lt}=\left [ \begin{array}{ccccccc}I &\quad 0 &\quad \cdots &\quad 0\\ I &\quad I &\quad \cdots &\quad 0 \\ \vdots &\quad \vdots &\quad \ddots &\quad \vdots \\ I &\quad I &\quad \cdots &\quad I \end{array}\right ]
\end{align}
The increment on the control input constraint as follows
\begin{equation}
         \Delta U_{\min } \leq  \Delta U_k \leq\Delta  U_{\max }
 \end{equation}
where $\Delta U_{\min } $,$\Delta  U_{\max }$, $U_{\min }$, and $U_{\max }$ are the lower and upper bounds of the input vectors.
We can define the cost function 
\begin{equation}
    \begin{array}{rl}
         {\textbf{min}}&\Delta U_k^TW\Delta U_k+c^T\Delta U_k  \\
         \textbf{subject to}& E\Delta U_k\leq b
    \end{array}
\end{equation}
where
\begin{align*}
    W=&(G_k^TQG_k+R),\\
    c=&2(F_k + g_k+d_k)^TQG_k \\
    E=&\left[
    \begin{array}{c}
         -I_{lt} \\
          I_{lt}\\
         -I_1\\
          I_1
    \end{array}
    \right]\quad I_1=I_{N_c\times N_c}\\
    b=&\left[
    \begin{array}{c}
         -U_{min}+u(k-1|k)1_{N_c}  \\
          U_{max}+u(k-1|k)1_{N_c} \\
          -\Delta U_{min}\\
           \Delta U_{max}
    \end{array}
    \right]
\end{align*}

\subsection{NMPC with Unknown Target Accelerations}
We can define the quadratic cost function for the same performance objective  (\ref{eq:prediction ss}) as follows\cite{2015PDNN,2020NMPC}:
\begin{equation}
    \begin{array}{rl}
         \text{min}&\left[\begin{array}{c}
              \Delta U_k  \\ d_k 
         \end{array}\right]^TW\left[\begin{array}{c}
              \Delta U_k  \\ d_k 
         \end{array}\right]+c^T\left[\begin{array}{c}
              \Delta U_k  \\ d_k 
         \end{array}\right]  \\
         \text{subject to}& E\left[\begin{array}{c}
              \Delta U_k  \\ d_k 
         \end{array}\right]\leq b
    \end{array}
\end{equation}
where
\begin{align*}
    W=\left[\begin{array}{cc}
         G_k^TQG_k+R&G^TQ  \\
         QG_k&Q 
    \end{array}\right],\quad c=\left[\begin{array}{cc}
         2G_k^TQ(F_k+g_k)  \\
         2Q(F_k+g_k) 
    \end{array}\right]\\
   E=\left[
    \begin{array}{cc}
         -I_{lt}&0_1 \\
          I_{lt}&0_1\\
         -I_1&0_1\\
          I_1&0_1\\
          0_2&-I_2\\
          0_2&I_2
    \end{array}
    \right]
    b=\left[
    \begin{array}{c}
         -U_{min}+u(k-1|k)1_{N_c}  \\
          U_{max}+u(k-1|k)1_{N_c} \\
          -\Delta U_{min}\\
           \Delta U_{max}\\
           -d_{min}\\
           d_{max}
    \end{array}
    \right]\\
    0_1=0_{N_p\times N_pn_x},0_2=0_{N_pn_x\times N_p}, I_2=I_{N_pn_x}
\end{align*}

% %%%%%%%%%%%%%%%%%%%%%%% Network Design %%%%%%%%%%%%%%%%
\section{PREDICTION ALGORITHM AND NETWORK DESIGN}
Predicting target motion is a challenging task due to highly nonlinear dynamics and large maneuver envelope of the aircraft. Furthermore, trajectories are further complicated by decisions executed by the pilot. That being said, many attack trajectories follow similar patterns \cite{lotter2017design}. Hence it is possible to either construct or collect data from radar to create maneuver library for such attacks. Then the trajectory prediction problem can be treated as a regular time series analysis problem on target trajectory. In this section, different types of neural network architectures for trajectory prediction task is discussed and evaluated.
%%%%%%%%%%%%%%%%%%%%%%%%%%%%%%%%%%%%%%%%%%%%%%%%%%%
%%%%%%%%%%%%%%%%%%%%%%%%%%% 
\subsection{Structured Attacks and Training Data}
\begin{figure*}[t]
    \centering
    \includegraphics[width=0.7\textwidth]{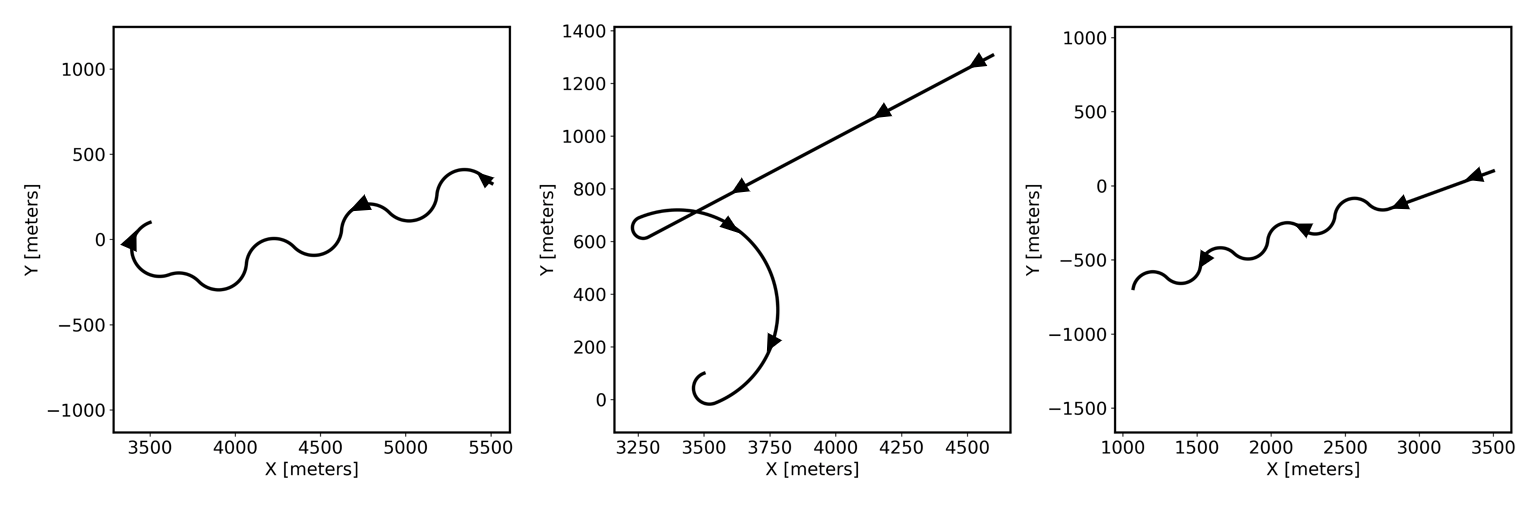}
    \caption{Simulation Scenarios; Combination of left, right coordinated turns, level flight and weave maneuver. Arrows indicate the flow of time.}
    \label{fig:flightdata}
\end{figure*}

To train and demonstrate prediction and interception method, a variety of interception scenarios are simulated. The maneuver parameters for each maneuver are selected randomly under the limits. In addition, target can switch one maneuver to another with equal probability. This allows us to simulate more complex structured attack patterns. Examples of the flight data can be seen on Fig. \ref{fig:flightdata}. For instance, the left side of the Fig. \ref{fig:flightdata} shows that target executes weave and coordinated turn maneuvers. The other sub-figures show that the target starts with level flight and executes high acceleration turn on the middle and weave maneuver at the right sub-figure.

A fixed origin frame of reference is used. The position and velocity vectors are obtained with respect to the origin. Input of the model can be written as;
\begin{align*}
    \text{X} &= [\text{x}_{t-1}, \text{x}_{t-2}, \text{x}_{t-3},....,\text{x}_{t-n}] \\
    \text{x}_t &= [x^t, y^t, v_x^t, v_y^t]
\end{align*}
\text{X} vector contains positions and velocities from $t-n$ to $t$, which $n$ is the observation history size. Min-max scaler normalization is used to fit the data into $[0, 1]$ interval since the target's position, velocity, and acceleration values vary differently. Gaussian noise is added to imitate the radar measurements.
% \begin{figure}[H]
%     \centering
%     \includegraphics[width=0.5\textwidth]{lstmfigure/Noisy.png}
%     \caption{Simulation data for target trajectories with/without noise}
%     \label{fig:noisy}
% \end{figure}
A total of 500 simulations with random initial positions and 30 seconds of flight time are conducted. Next, the data are randomly sampled from flights in the form of $X$. The data are divided such that \% 60 is used for training, \%20 for validation and \%20 for test. During the simulation studies, it has been observed that at least 2 seconds of observation history is essential to obtain good predictions. %At this point, we suggest to train networks with very low and very high observation history and find the optimum point in between. Interested reader may also check hyper parameter optimization tools for neural networks.
\subsection{Network Design for Trajectory Prediction}
%%%%%%%%%%%%%%%%%%%%%%%%%%%%%%%%%%%%%%
There are plenty of deep neural network architectures that consist of Convolutional Neural Networks (CNNs), Recurrent Neural Networks (RNNs), Long-Short Term Memory Networks(LSTMs) for time-series analysis. The CNN architecture is mostly used for spatial inference from high dimensional data such as an image, and it can also be used for one-dimensional long-term dependent data \cite{LSTM5}. It is well known that  LSTM and RNN models have more suitable structures to process time series data. Since the input and output size are different,an encoder-decoder architecture is needed to deal with sequence to sequence prediction\cite{LSTM3,LSTM2}.Moreover, this type of architecture is capable of generalizing the solution and remove input noise.
\begin{figure}[H]
    \centering
    \includegraphics[width=0.4\textwidth]{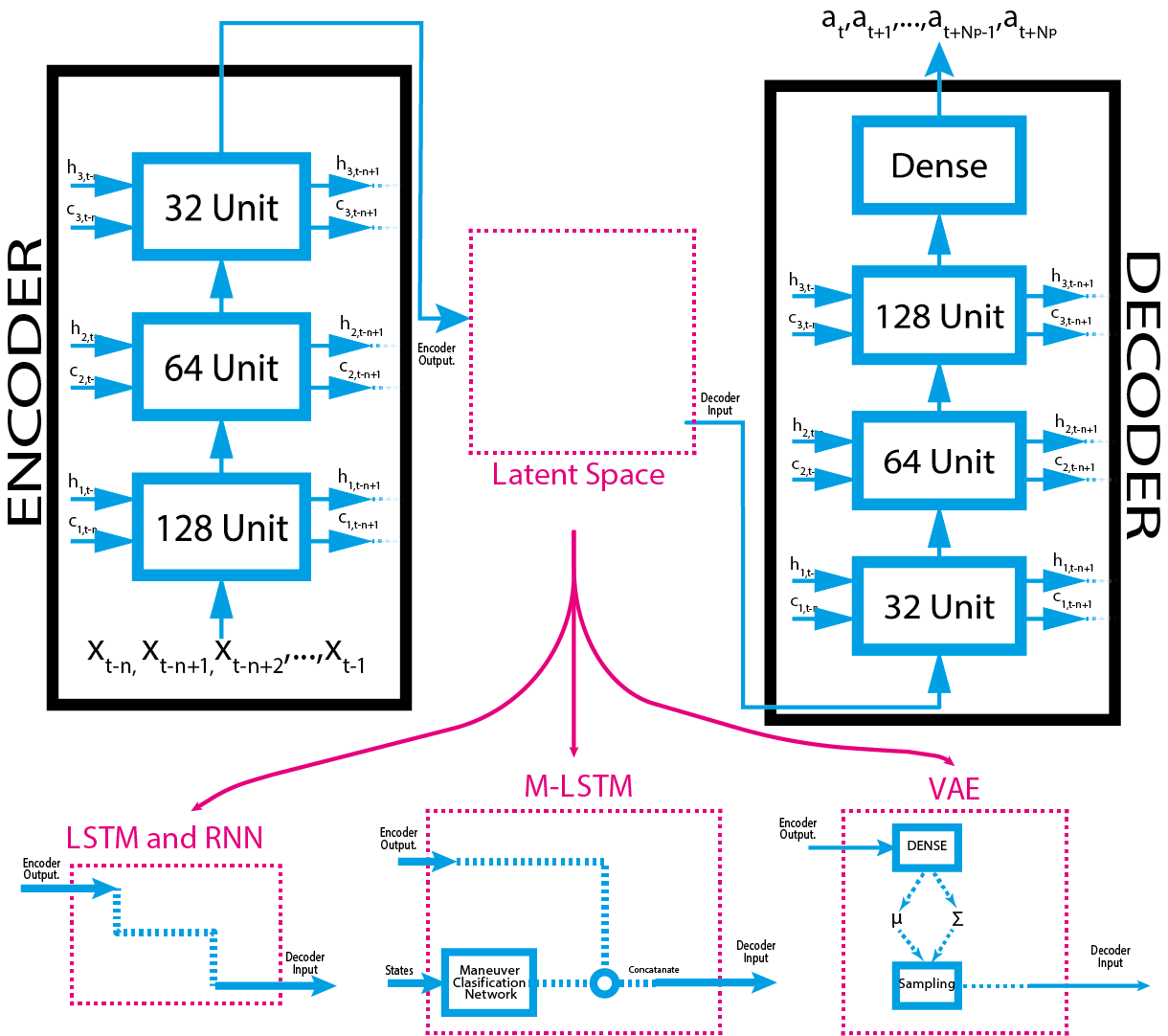}
    \caption{Encoder-Decoder architectures for trajectory prediction.}
    \label{fig:architecture}
\end{figure}
Our preliminary studies about which kind of neural network model performs more precise acceleration predictions give rise to implement LSTM and RNN models for prediction. So, four kinds of recurrent neural network architectures such as LSTM, RNN, MLST and variational autoencoders (VAEs) \cite{LSTM4} are compared with each other. Fig. \ref{fig:architecture} shows the studied architectures constructed over encoder-decoder. The first LSTM and RNN models are vanilla encoder-decoder which directly connect encoder outputs to decoder inputs at the latent space. On the other hand, M-LSTM adds a maneuver classification feature into encoder outputs then sends it to decoder input. It is inspired M-LSTM model from \cite{LSTM2} since an LSTM encoder-decoder architecture with maneuver classification improves the accuracy of position prediction. In addition to the models above, variational autoencoders derive a Gauss distribution and then take a sample from it in the latent space.
\subsection{Implementation Details}
Parameters for the prediction model are tuned according to simulation results. It is observed that adding extra layers increases computation time rather than improving the accuracy of the model. However, an increasing number of neurons in a single layer may improve the overall model accuracy. \\
Activation functions at the last layers of the classification part in the M-LSTM are softmax. In order to prevent over-fitting, dropout method is employed. For recurrent layers 20\% and for dense layers 10\% dropout is implemented. Simulation studies show that Adagrad optimization method which changes learning rates according to parameter update frequency is a good fit for this problem. 
The number of trainable parameters varies in different architectures. The strategy was to select first encoder layer as the biggest and then decreasing and increasing the layer through the encoder decoder layers.
% For instance, small size of the encoder architecture contains 2 layers and neuron size with 64 and 32. The large parameter models contain 128, 64, 32 neuron encoder layers and 32, 64, 128 decoder layers, respectively. 
No further improvement is obtained above 128 neurons.
\subsection{The Prediction Results}
In order to certify the effectiveness of the proposed model, SimpleRNN, LSTM, M-LSTM and VAE models are compared in simulation. The training and test accuracy are measured with Mean Square Error (MSE). The results can be seen at Table \ref{tab:traintest}. The acceleration limits, $[-25, 25]$, were normalized to $[0, 1]$ interval. Therefore, the MSE losses must be examined accordingly.
\begin{table}[H]
\centering
\caption{Training and Test Losses of the Models}
\label{tab:traintest}
\resizebox{0.5\textwidth}{!}{%
\begin{tabular}{|lllllllll|}
\hline
 & \multicolumn{2}{c}{SimpleRNN} & \multicolumn{2}{c}{LSTM} & \multicolumn{2}{c}{M-LSTM} & \multicolumn{2}{c|}{VAE} \\
Number of Parameters & \multicolumn{1}{c}{Train} & \multicolumn{1}{c}{Test} & \multicolumn{1}{c}{Train} & \multicolumn{1}{c}{Test} & \multicolumn{1}{c}{Train} & \multicolumn{1}{c}{Test} & \multicolumn{1}{c}{Train} & \multicolumn{1}{c|}{Test} \\ \hline
Small & 0.0304 & 0.0373 & 0.0121 & 0.0191 & 0.0175 & 0.0156 & 0.0653 & 0.0648 \\
Medium & 0.0258 & 0.0328 & 0.0169 & 0.0152 & 0.0158 & 0.0142 & 0.0651 & 0.0641 \\
Large & 0.0376 & 0.0386 & 0.0162 & 0.0096 & 0.0148 & 0.0091 & 0.0648 & 0.0631 \\ \hline
\end{tabular}%
}
\end{table}
The result show that LSTM and M-LSTM architectures are superior to SimpleRNN and VAE. In addition, M-LSTM is slightly better than LSTM architecture. Therefore, in the presence of maneuver label information, more accurate models can be used. All models also perform well on test data, which shows that models are capable of generalizing to maneuvers not used in training.  
% All model perform better on test data. That means models actually learn instead of memorizing.

\begin{table}[H]
\centering
\caption{Example Prediction Results on Test Data}
\label{tab:predictionResults}
\resizebox{0.45\textwidth}{!}{%
\begin{tabular}{|ccccccccc|}
\hline
\multicolumn{1}{|l}{} & \multicolumn{2}{c}{10 Horizon} & \multicolumn{2}{c}{10 Horizon} & \multicolumn{2}{c}{20 Horizon} & \multicolumn{2}{c|}{20 Horizon} \\
Time Step & Real & Prediction & Real & Prediction & Real & Prediction & Real & Prediction \\ \hline
\multicolumn{1}{|c|}{k} & \multicolumn{1}{c|}{-16.21} & \multicolumn{1}{c|}{-16.88} & \multicolumn{1}{c|}{-24.67} & \multicolumn{1}{c|}{-25} & \multicolumn{1}{c|}{14.829} & \multicolumn{1}{c|}{14.572} & \multicolumn{1}{c|}{15.182} & 14.749 \\ \hline
\multicolumn{1}{|c|}{k+1} & \multicolumn{1}{c|}{-16.21} & \multicolumn{1}{c|}{-15.74} & \multicolumn{1}{c|}{-24.67} & \multicolumn{1}{c|}{-25} & \multicolumn{1}{c|}{14.829} & \multicolumn{1}{c|}{14.643} & \multicolumn{1}{c|}{15.182} & 13.991 \\ \hline
\multicolumn{1}{|c|}{.} & \multicolumn{1}{c|}{16.21} & \multicolumn{1}{c|}{-8.67} & \multicolumn{1}{c|}{-24.67} & \multicolumn{1}{c|}{-25} & \multicolumn{1}{c|}{.} & \multicolumn{1}{c|}{.} & \multicolumn{1}{c|}{.} & . \\ \hline
\multicolumn{1}{|c|}{.} & \multicolumn{1}{c|}{16.21} & \multicolumn{1}{c|}{12.18} & \multicolumn{1}{c|}{-24.67} & \multicolumn{1}{c|}{-11.08} & \multicolumn{1}{c|}{.} & \multicolumn{1}{c|}{.} & \multicolumn{1}{c|}{15.182} & 12.97 \\ \hline
\multicolumn{1}{|c|}{.} & \multicolumn{1}{c|}{16.21} & \multicolumn{1}{c|}{19.25} & \multicolumn{1}{c|}{24.67} & \multicolumn{1}{c|}{20.57} & \multicolumn{1}{c|}{.} & \multicolumn{1}{c|}{.} & \multicolumn{1}{c|}{15.182} & 9.227 \\ \hline
\multicolumn{1}{|c|}{.} & \multicolumn{1}{c|}{16.21} & \multicolumn{1}{c|}{17.84} & \multicolumn{1}{c|}{24.67} & \multicolumn{1}{c|}{25.11} & \multicolumn{1}{c|}{14.829} & \multicolumn{1}{c|}{13.624} & \multicolumn{1}{c|}{15.182} & -0.084 \\ \hline
\multicolumn{1}{|c|}{.} & \multicolumn{1}{c|}{16.21} & \multicolumn{1}{c|}{17.54} & \multicolumn{1}{c|}{24.67} & \multicolumn{1}{c|}{23.83} & \multicolumn{1}{c|}{14.829} & \multicolumn{1}{c|}{13.61} & \multicolumn{1}{c|}{.} & . \\ \hline
\multicolumn{1}{|c|}{.} & \multicolumn{1}{c|}{16.21} & \multicolumn{1}{c|}{17.26} & \multicolumn{1}{c|}{24.67} & \multicolumn{1}{c|}{23.91} & \multicolumn{1}{c|}{14.829} & \multicolumn{1}{c|}{13.578} & \multicolumn{1}{c|}{-15.182} & -15.699 \\ \hline
\multicolumn{1}{|c|}{k+n} & \multicolumn{1}{c|}{16.21} & \multicolumn{1}{c|}{17.22} & \multicolumn{1}{c|}{24.67} & \multicolumn{1}{c|}{24.19} & \multicolumn{1}{c|}{14.829} & \multicolumn{1}{c|}{13.561} & \multicolumn{1}{c|}{-15.182} & -16.032 \\ \hline
\end{tabular}%
}
\end{table}
Several acceleration prediction results can be seen at Table \ref{tab:predictionResults}. The results show that the model can predict target acceleration with high accuracy on a large set of agile maneuvers. It should be noted that the model also captures and predicts switches between different maneuvers types and associated accelerations. This property provides our proposed method NMPC-TAP a considerable edge against the alternative approaches.
\section{SIMULATION RESULTS}
In this section, simulation studies are conducted by comparing proposed NMPC-TAP algorithm with aforementioned methods from literature such as PN, APN and NMPC. Trajectory predictions governed by the M-LSTM from table \ref{tab:traintest}. To compare these methods, 100 Monte-Carlo simulation are run for each guidance algorithm with $ 5\% $ noise. Simulation parameters of NMPC-TAP are given by Table \ref{table:simulation param.}.

%     g=9.81 [m/s^2],\Delta t=0.02 [s],u_{max}=25g ,u_{min}=-u_{max}, \Delta u_{max}=0.025u_{max},\Delta u_{min}=-\Delta u_{max},d_{max}=8g\Delta t [m/s^2],d_{min}=-d_7max},V_M =150 [m/s],V_T=100 [m/s], Q=diag[q,q,\dots,q] and q=[0,0,0,100] 

\begin{table}[H]
	\begin{center}
		\caption{Simulation parameters}
		\label{table:simulation param.}
		\resizebox{0.4\textwidth}{!}{
		\begin{tabular}{ |c|c||c|c| }
			\hline
			$\Delta t$&0.02 [s]&$d_{max}$ & $8g\times\Delta t [m/s^2]$\\
			\hline
			g	&9.81 $[m/s^2]$&$d_{min}$& $-d_{max}$\\
			\hline
			$u_{max}$&25g &$\Delta u_{max}$    &0.025$u_{max}$\\
			\hline
			$u_{min}$&$-u_{max}$&$\Delta u_{min}$    &$-\Delta u_{max}$\\
			\hline
			q & [0,0,0,100]&R&I \\
			\hline
			$\theta_{M,0}$&$0^{\circ}$&$\theta_{T,0}$ &$190^{\circ}$\\
			\hline
			$x_{M,0}$ &0&$x_{T,0}$ &1000\\
			\hline
			$y_{M,0}$&0&$y_{T,0}$&1000\\
			\hline
			$V_M$ &150&$V_T$&100\\
			\hline
			\multicolumn{4}{|c|}{$N^\prime=3$}\\
			\hline
		\end{tabular} }
	\end{center}
\end{table} 

\begin{figure}[h]
    \centering
    \includegraphics[scale=0.27]{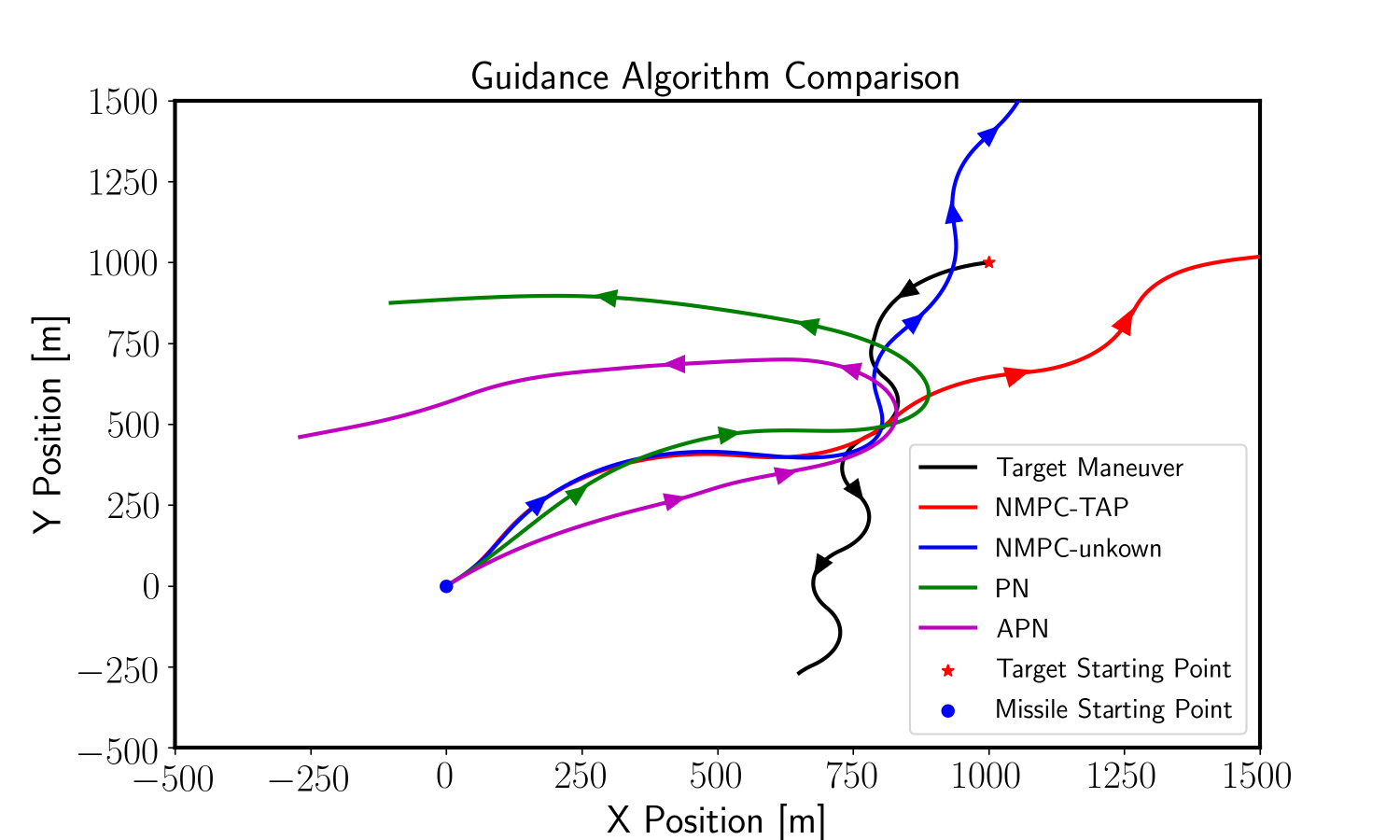}
    \caption{Missile trajectories generated by each guidance algorithms during missile-target interception.}
    \label{fig:trajectories}
\end{figure}

\begin{figure}[h]
    \centering
    \includegraphics[scale=0.27]{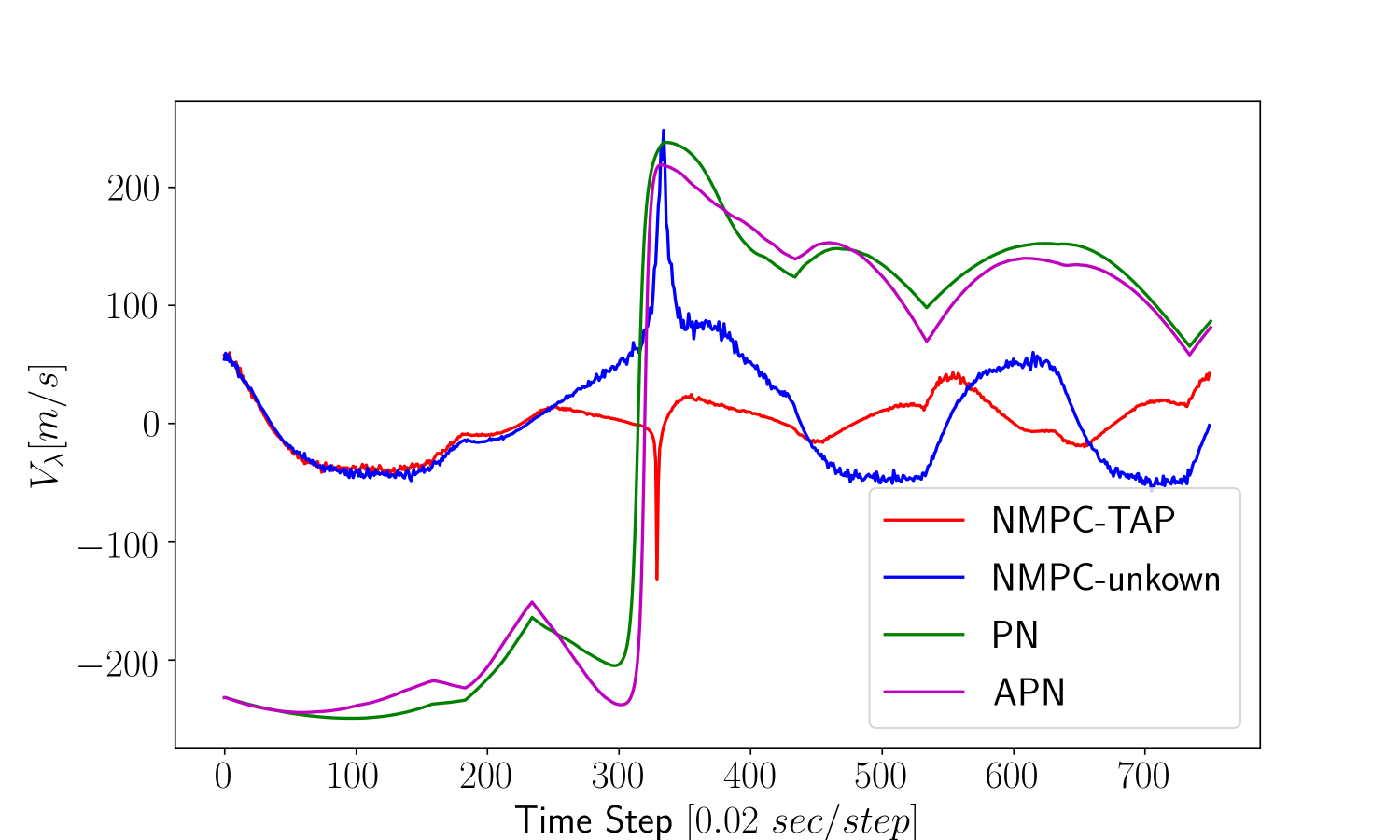}
    \caption{$V_\lambda$ histories during missile-target interception.}
    \label{fig:V_lambda }
\end{figure}

\begin{figure}[h]
    \centering
    \includegraphics[scale=0.27]{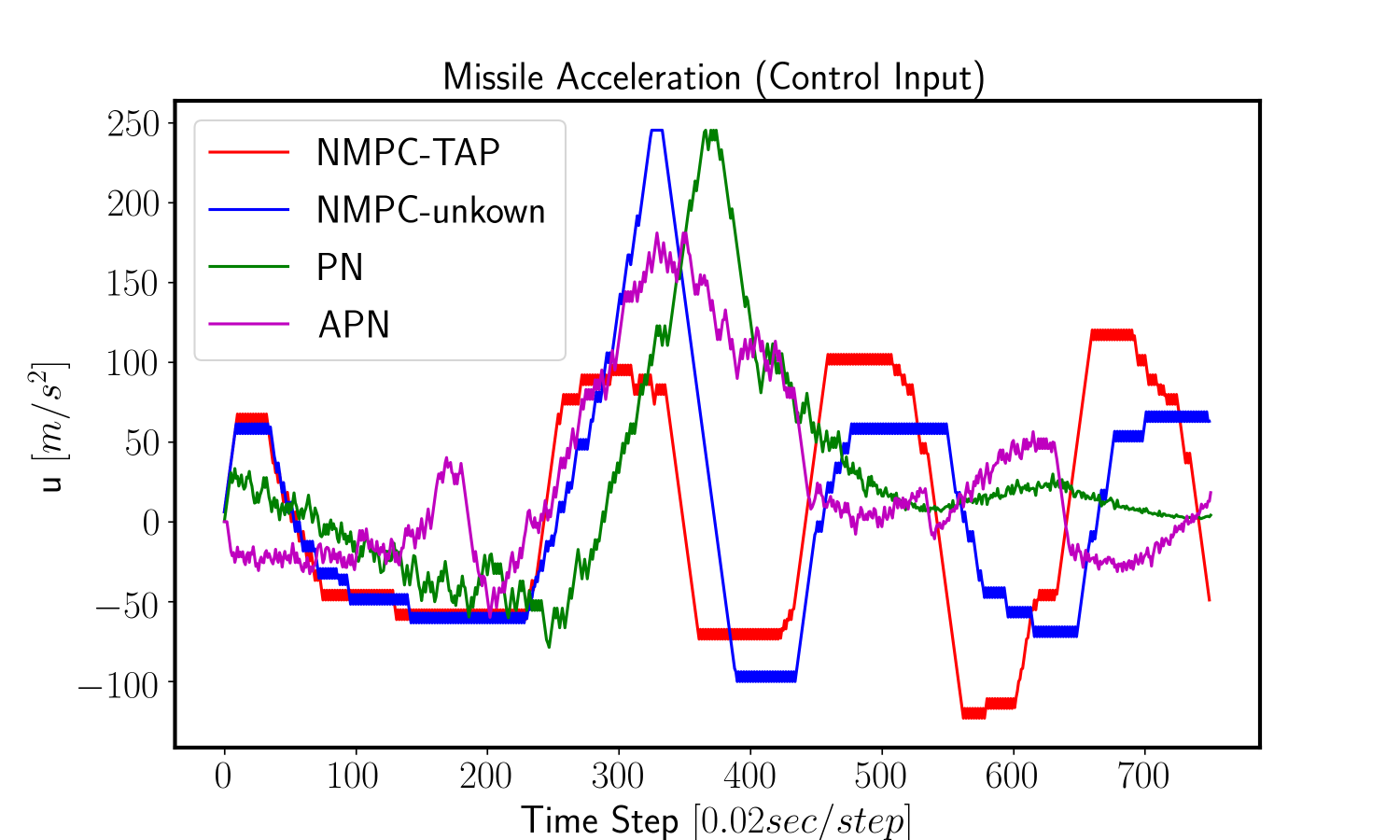}
    \caption{Missile acceleration histories during missile-target interception.}
    \label{fig:input_acc}
\end{figure}

Here $Q=diag([q,\dots,q]) $ and $q \in\Re^{n_x\times1}$, $N^\prime$ is navigation constant for PN and APN. The NMPC-TAP cost function Eq. (\ref{eq:cost}) is chosen to penalize the $V_\lambda$ and input rate. An agile maneuvering target is used for simulation purposes. The maneuver is assumed that the target begins with left coordinated turn with $4g$ and $70^{\circ}$ then makes $100m$ level flight and weave maneuver with $8g$ and $4s$ period. Interception performance of each guidance law against aforementioned agile maneuvering target is shown in Fig. \ref{fig:trajectories}. 
It is apparently seen that each approach achieves target interception. However the interception performance can not be evaluated objectively by only analyzing X, Y trajectories.
Then, relative velocity component perpendicular to the LOS, $V_\lambda$ is given by Fig. \ref{fig:V_lambda }, since the quadratic cost Eq. (\ref{eq:cost}) in NMPC based approaches primarily aims to penalize $V_\lambda$.
NMPC unknown and NMPC-TAP has superior $V_\lambda$ responses. $V_\lambda$ is being enforced to zero much more earlier than PN and APN laws with a navigation constant equal to 3. Moreover, NMPC-TAP provides smaller magnitudes along whole simulation history. Fig.\ref{fig:input_acc} shows the control input histories.
Benefits of employing target acceleration predictions in optimization problem is revealed by control inputs. NMPC-TAP changes direction of control input few steps earlier than NMPC-unknown. In addition, control input constrains are not violated by both NMPC based approaches.
\begin{table}[h]
    \caption{Results obtained by MC Simulations.}
    \label{table:guid. comp.}
    \begin{center}
    \resizebox{0.9\columnwidth}{!}{
        \begin{tabular}{|c|c|c|c|c|}

        \hline
         & $MD_{mean}$ (m)& $MD_{STD}$ (m) & IT (s) & AIR $(m/s^2)$\\
        \hline
        APN & 18.69 & 22.75 &6.52 &5.21\\
        \hline
        PN & 26.378 & 7.72 &6.52 &5.3\\
        \hline
        \multicolumn{5}{|c|}{Results for $N_p=10$}\\
        \hline
        NMPC-TAP & 7.2 & 12.91 &6.54 &5.64\\
        \hline
        NMPC & 9.2 & 17.49 &6.67 &4.76\\
        \hline
        \multicolumn{5}{|c|}{Results for $N_p=30$}\\
        \hline
        NMPC-TAP & 6.58 & 8.80 &6.56 &6.06\\
        \hline
        NMPC & 11.22 & 6.19 &6.57 &6.06\\
        \hline
        \multicolumn{5}{|c|}{Results for $N_p=40$}\\
        \hline
        NMPC-TAP & 3.57 & 2.70 &6.62 &6.025\\
        \hline
        NMPC & 14.58 & 6.05 &6.73 &5.96\\
        \hline
        \end{tabular}
        }
    \end{center}
\end{table}
In order to increase reliability of comparisons, results are averaged over 100 Monte-Carlo (MC) simulations with the values of $N_p=10$, $N_p=30$ and $N_p=40$. It is also assumed that $N_p=N_c$ for each simulations. The results are evaluated in terms of miss distance, interception time and input rate. Table\ref{table:guid. comp.} compares PN, APN, NMPC and NMPC-TAP algoritms as given above.\\
Here, $MD_{mean}$ denotes mean of miss distances, $MD_{STD}$ represents standard deviation of miss distances, IT stands for interception time and AIR presents average absolute input rate. Increasing the dimension of prediction horizon is efficiently decreasing the $MD_{mean}$  values. Since the target is performing agile escape maneuvers, LSTM predictions provide invaluable future information along the prediction horizon. NMPC-TAP outperforms other guidance algorithms as long as the prediction horizon is sufficiently long. Despite the fact that NMPC-TAP presents much lower $MD_{mean}$ values, limitations on the allowable control signal is not sacrificed.  

% %%%%%%%%%%%%%%%%%%%%% Conclusions %%%%%%%%%%%%%%%
\section{CONCLUSIONS}
A novel nonlinear model predictive guidance that uses target acceleration prediction provided by an LSTM NN is proposed in this work. Target acceleration predictions are directly included in the NMPC scheme for propagating the interception problem's states. The NMPC formulates the interception problem as a QP problem in order to find the sub-optimal missile acceleration command. Simulations, which are conducted against an agile target to evaluate the performance and effectiveness of the proposed method, demonstrate that the desired guidance method is superior when compared to the others.

\section*{ACKNOWLEDGMENT}
This work is supported by the ITU BAP grant no: MOA-2019-42321.

\bibliographystyle{IEEEtran}
\bibliography{IEEEabrv,references}

\begin{thebibliography}{10}
\providecommand{\url}[1]{#1}
\csname url@rmstyle\endcsname
\providecommand{\newblock}{\relax}
\providecommand{\bibinfo}[2]{#2}
\providecommand\BIBentrySTDinterwordspacing{\spaceskip=0pt\relax}
\providecommand\BIBentryALTinterwordstretchfactor{4}
\providecommand\BIBentryALTinterwordspacing{\spaceskip=\fontdimen2\font plus
\BIBentryALTinterwordstretchfactor\fontdimen3\font minus
  \fontdimen4\font\relax}
\providecommand\BIBforeignlanguage[2]{{%
\expandafter\ifx\csname l@#1\endcsname\relax
\typeout{** WARNING: IEEEtran.bst: No hyphenation pattern has been}%
\typeout{** loaded for the language `#1'. Using the pattern for}%
\typeout{** the default language instead.}%
\else
\language=\csname l@#1\endcsname
\fi
#2}}

\bibitem{zarchan2012tactical}
P.~Zarchan, \emph{Tactical and strategic missile guidance}.\hskip 1em plus
  0.5em minus 0.4em\relax American Institute of Aeronautics and Astronautics,
  Inc., 2012.

\bibitem{garber1968optimum}
V.~Garber, ``Optimum intercept laws for accelerating targets.'' \emph{AIAA
  Journal}, vol.~6, no.~11, pp. 2196--2198, 1968.

\bibitem{ghosh2013capturability}
S.~Ghosh, D.~Ghose, and S.~Raha, ``Capturability of augmented proportional
  navigation (apn) guidance with nonlinear engagement dynamics,'' in \emph{2013
  American Control Conference}.\hskip 1em plus 0.5em minus 0.4em\relax IEEE,
  2013, pp. 7--12.

\bibitem{ghosh2014capturability}
------, ``Capturability of augmented pure proportional navigation guidance
  against time-varying target maneuvers,'' \emph{Journal of Guidance, Control,
  and Dynamics}, vol.~37, no.~5, pp. 1446--1461, 2014.

\bibitem{zhurbal2011effect}
A.~Zhurbal and M.~Idan, ``Effect of estimation on the performance of an
  integrated missile guidance and control system,'' \emph{IEEE Transactions on
  Aerospace and Electronic systems}, vol.~47, no.~4, pp. 2690--2708, 2011.

\bibitem{prabhakar2013predictive}
N.~Prabhakar, I.~D. Kumar, S.~K. Tata, and V.~Vaithiyanathan, ``A predictive
  explicit guidance scheme for ballistic missiles.'' \emph{Defence science
  journal}, 2013.

\bibitem{ure2016predictive}
N.~K. Ure and G.~Inalhan, ``Predictive missile guidance for agile maneuvering
  targets with stochastic hybrid dynamics,'' in \emph{2016 IEEE Aerospace
  Conference}.\hskip 1em plus 0.5em minus 0.4em\relax IEEE, 2016, pp. 1--9.

\bibitem{akcal2017predictive}
M.~U. Akcal and N.~K. Ure, ``Predictive missile guidance with online trajectory
  learning.'' \emph{Defence Science Journal}, vol.~67, no.~3, 2017.

\bibitem{best2000predictive}
R.~Best and J.~Norton, ``Predictive missile guidance,'' \emph{Journal of
  Guidance, Control, and Dynamics}, vol.~23, no.~3, pp. 539--546, 2000.

\bibitem{shima2002efficient}
T.~Shima, Y.~Oshman, and J.~Shinar, ``Efficient multiple model adaptive
  estimation in ballistic missile interception scenarios,'' \emph{Journal of
  Guidance, Control, and Dynamics}, vol.~25, no.~4, pp. 667--675, 2002.

\bibitem{dionne2006predictive}
D.~Dionne, H.~Michalska, and C.~A. Rabbath, ``A predictive guidance law with
  uncertain information about the target state,'' in \emph{2006 American
  Control Conference}.\hskip 1em plus 0.5em minus 0.4em\relax IEEE, 2006, pp.
  6--pp.

\bibitem{nobahari2016heuristic}
H.~Nobahari and A.~Haeri, ``A heuristic predictive los guidance law based on
  trajectory learning, ant colony optimization and tabu search,'' in \emph{2016
  6th IEEE International Conference on Control System, Computing and
  Engineering (ICCSCE)}.\hskip 1em plus 0.5em minus 0.4em\relax IEEE, 2016, pp.
  163--168.

\bibitem{MPCsurvey_2017_AIAA}
U.~Eren, A.~Prach, B.~B. Ko{\c{c}}er, S.~V. Rakovi{\'c}, E.~Kayacan, and
  B.~A{\c{c}}{\i}kme{\c{s}}e, ``Model predictive control in aerospace systems:
  Current state and opportunities,'' \emph{Journal of Guidance, Control, and
  Dynamics}, vol.~40, no.~7, pp. 1541--1566, 2017.

\bibitem{2015PDNN}
Z.~{Li}, Y.~{Xia}, C.~{Su}, J.~{Deng}, J.~{Fu}, and W.~{He}, ``Missile guidance
  law based on robust model predictive control using neural-network
  optimization,'' \emph{IEEE Transactions on Neural Networks and Learning
  Systems}, vol.~26, no.~8, pp. 1803--1809, 2015.

\bibitem{bachtiar2017nonlinear}
V.~Bachtiar, C.~Manzie, and E.~C. Kerrigan, ``Nonlinear model-predictive
  integrated missile control and its multiobjective tuning,'' \emph{Journal of
  Guidance, Control, and Dynamics}, vol.~40, no.~11, pp. 2961--2970, 2017.

\bibitem{kumar2020nonlinear}
P.~Kumar, S.~Sonkar, A.~Ghosh, and D.~Philip, ``Nonlinear model-predictive
  integrated guidance and control scheme applied for missile-on-missile
  interception,'' in \emph{2020 International Conference on Emerging Smart
  Computing and Informatics (ESCI)}.\hskip 1em plus 0.5em minus 0.4em\relax
  IEEE, 2020, pp. 318--324.

\bibitem{2020NMPC}
D.~Bhattacharjee, A.~Chakravarthy, and K.~Subbarao, ``Nonlinear model
  predictive control based missile guidance for target interception,'' in
  \emph{AIAA Scitech 2020 Forum}, 2020, p. 0865.

\bibitem{lotter2017design}
D.~P. Lotter, ``Design of a weapon assignment subsystem within a ground-based
  air defence environment,'' Ph.D. dissertation, Stellenbosch: Stellenbosch
  University, 2017.

\bibitem{2001model}
D.~Lianos, Y.~Shtessel, and I.~Shkolnikov, ``Integrated guidance-control system
  of a homing interceptor-sliding mode approach,'' in \emph{AIAA Guidance,
  Navigation, and Control Conference and Exhibit}, 2001, p. 4218.

\bibitem{2006receding}
W.~H. Kwon and S.~H. Han, \emph{Receding horizon control: model predictive
  control for state models}.\hskip 1em plus 0.5em minus 0.4em\relax Springer
  Science \& Business Media, 2006.

\bibitem{LSTM5}
B.~Zhao, H.~Lu, S.~Chen, J.~Liu, and D.~Wu, ``Convolutional neural networks for
  time series classification,'' \emph{Journal of Systems Engineering and
  Electronics}, vol.~28, pp. 162--169, 02 2017.

\bibitem{LSTM3}
I.~Sutskever, O.~Vinyals, and Q.~Le, ``Sequence to sequence learning with
  neural networks,'' \emph{Advances in Neural Information Processing Systems},
  vol.~4, 09 2014.

\bibitem{LSTM2}
N.~{Deo} and M.~M. {Trivedi}, ``Multi-modal trajectory prediction of
  surrounding vehicles with maneuver based lstms,'' in \emph{2018 IEEE
  Intelligent Vehicles Symposium (IV)}, 2018, pp. 1179--1184.

\bibitem{LSTM4}
D.~Kingma and M.~Welling, ``An introduction to variational autoencoders,''
  \emph{Foundations and Trends® in Machine Learning}, vol.~12, pp. 307--392,
  01 2019.

\end{thebibliography}

\end{document}